\documentclass[a4paper,10pt]{article}
\usepackage{graphicx} \usepackage{amsmath}

\title{Scaling Laws for Spreading of a Liquid Under Pressure}
\author{Soma Nag, Suparna Dutta and Sujata Tarafdar$^*$}

\begin{document}

\maketitle \begin{center}

\noindent
Condensed Matter Physics Research Centre, Physics Department\\ Jadavpur University, Kolkata 700032, India\\ \noindent
${^*}$ Corresponding author: Email: sujata$\_$tarafdar@hotmail.com,\\  Fax: +91 33 24138917, Phone:+913324146666(extn. 2760)
\end{center}

\begin{abstract} \noindent
We study squeeze flow of two different fluids (castor oil and ethylene glycol) between a pair of glass plates and a pair of perspex plates, under an applied load. The film thickness is found to vary with time as a power-law, where the exponent increases with load. After a certain time interval the area of fluid-solid contact saturates to a constant value. This saturation area, increases with load at different rates for different fluid-solid combinations.
\end{abstract}

\section{Introduction}
The spreading of a fluid on a solid substrate depends on the nature of the fluid as well as the substrate and also on other conditions such as temperature and humidity. Though this process is a common everyday experience, details of the spreading mechanism are still incompletely understood. A small droplet of a liquid, placed on a solid plate will take the shape of a section of a sphere. Under equilibrium, the well known Young-Laplace relation holds, involving an equilibrium angle of contact $\theta_e$. Under dynamic conditions however, while the fluid is spreading, the situation is more complicated. There are a number of excellent reviews summarizing studies on - the rate of spreading, the forces responsible and other interesting features \cite{degennes,bonn,engmann,asthana}. In the present study we look at a simple extension of the problem - what happens if the fluid is forced to spread beyond its equilibrium area, under an impressed force.

This is close to the `squeeze film' problem of chemical engineering\cite{engmann}, but usually the squeeze film with constant area  is considered, rather than the \textit{constant mass} set up we are interested in. Our set up is very simple. The fluid is sandwiched between two transparent horizontal  plates. A weight is placed on the upper plate. As the fluid spreads, it is photographed by a video camera, placed below the lower plate. 
Our observations cover two aspects. 

(i) The dynamics: where we
 find that under different loads the curves for film thickness $h$ vs. time show a power-law behavior, where the exponent increases with the load.

(ii) The statics: the area of contact of liquid and plate reaches a more-or-less constant value $A_{sat}$ after some time and the saturated 
area is plotted against load. The results presented here are for one water-soluble and one water-insoluble liquid on two different substrates - one low energy and the other with relatively higher energy. We try to find an explanation for the observed behavior in cases (i) and (ii). 

\section{Experimental}
The experimental fluids are castor oil and ethylene glycol, insoluble and soluble in water respectively. A constant volume $V_f$ of the fluid is measured with a micropipette and placed on the lower plate. The upper plate is then placed on the drop and a load of M kg. (M =  to 15 kg) is placed on it. The pair of plates are $\sim$ 1 cm thick, one pair is of float glass and the other perspex. A video camera, placed below the lower glass plate records the slow increase in area. The movies are analyzed using Image Pro-plus software. $V_f$ is taken as 20 $\mu l$ for ethylene glycol and $ 40 \mu l$ for castor oil. A larger volume of ethylene glycol spreads to too large areas for our set up and less castor oil does not spread appreciably, hence this choice.

\section{Results} We measure the area $A(t)$ and plot area vs. time curves for different loads with different fluid-solid combinations. Assuming $V_f$ to be constant the plate separation i.e. film thickness $h(t)$ is determined. Video frames are recorded normally at the rate of 4 frames per second until the area becomes more-or-less constant. However, for a few runs, we record the first 2-3 seconds at 10 frames per second to detect the initial plate velocity and its change precisely.

\begin{figure}[ht]
\begin{center}
\includegraphics[width=12.0cm, angle=0]{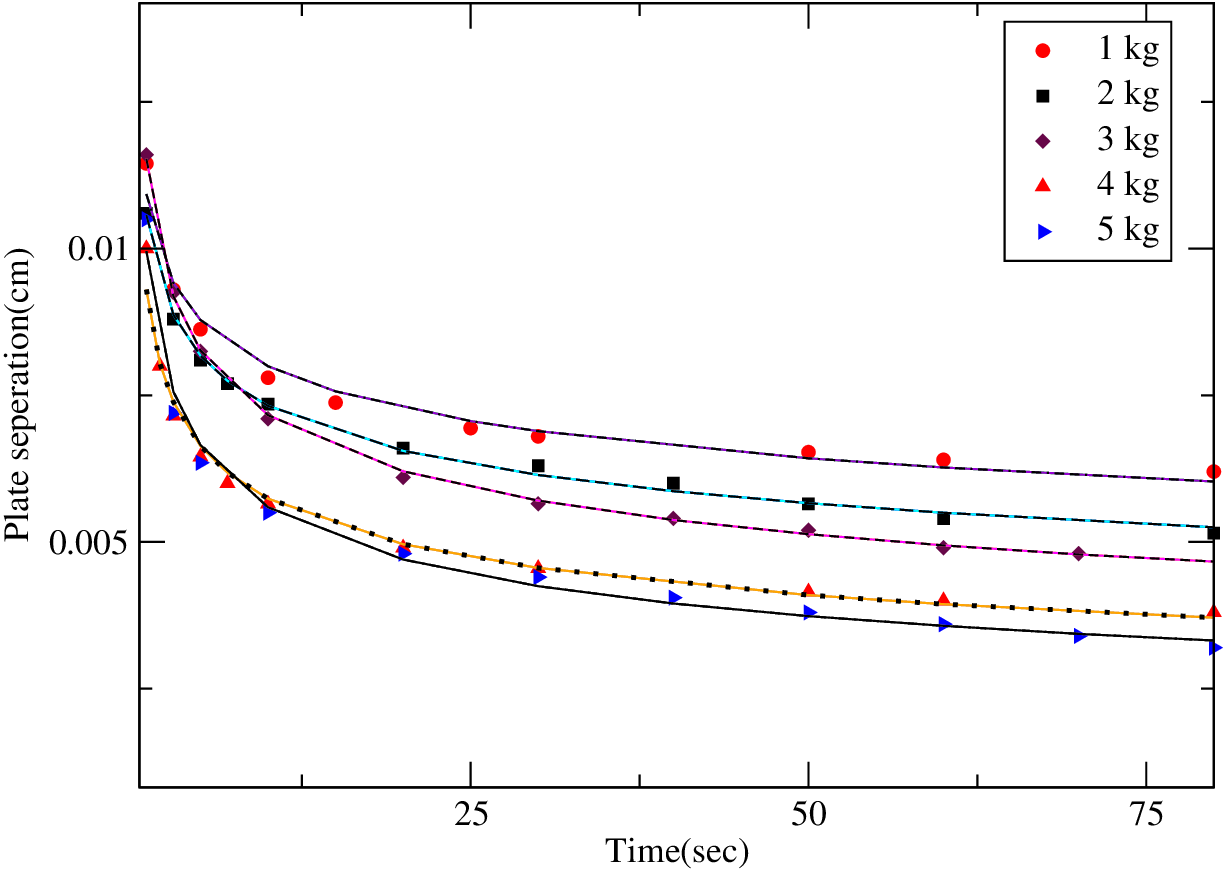}
\end{center}
\caption{Plate separation with time for castor oil on perspex for different loads. Symbols represent experimental points and lines are power-law fits to the data.
 } \label{htcp}
\end{figure}

$A(t)$ and hence $h(t)$ vary in time following a power-law and the exponent $m(M)$ is an increasing function of the load $M$. This is shown in figure(\ref{htcp}) for castor oil on perspex. 
The results give an excellent fit with the exponent varying between 0.14 to 0.25. These values are consistent with experiments for spreading without load \cite{degennes} reported to be around 0.2, result known as Tanner's law \cite{bonn,degennes}. For ethylene glycol the exponents cover a wider range, from 0.09 to 0.55, however the data points are more scattered here and exact values are less reliable. 

In general therefore, we have the relation
\begin{equation} A(t) \propto 1/{h(t)} \sim t^{m(M)} \end{equation} 

\begin{figure}[ht]
\begin{center}
\includegraphics[width=12.0cm, angle=0]{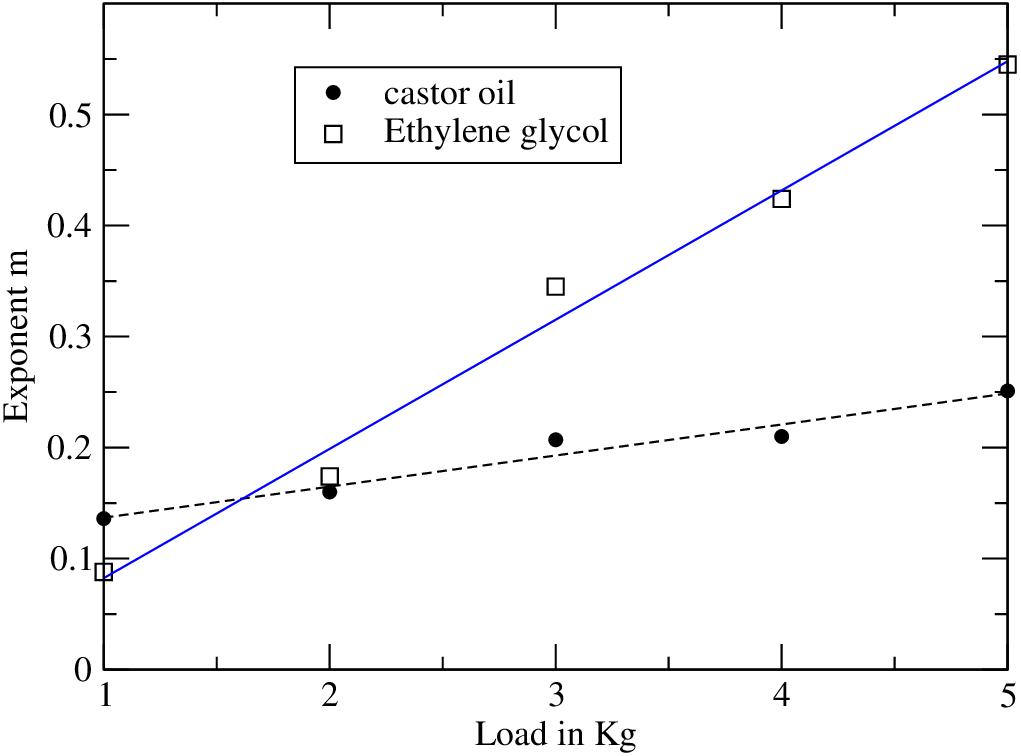}
\end{center}
\caption{Exponents for the power-law fits to the area vs. time data for ethylene glycol on glass and perspex are shown as function of load. Symbols are data points and lines, linear fits.
 } \label{exponents}
\end{figure}

\begin{figure}[ht]
\begin{center}
\includegraphics[width=12.0cm, angle=0]{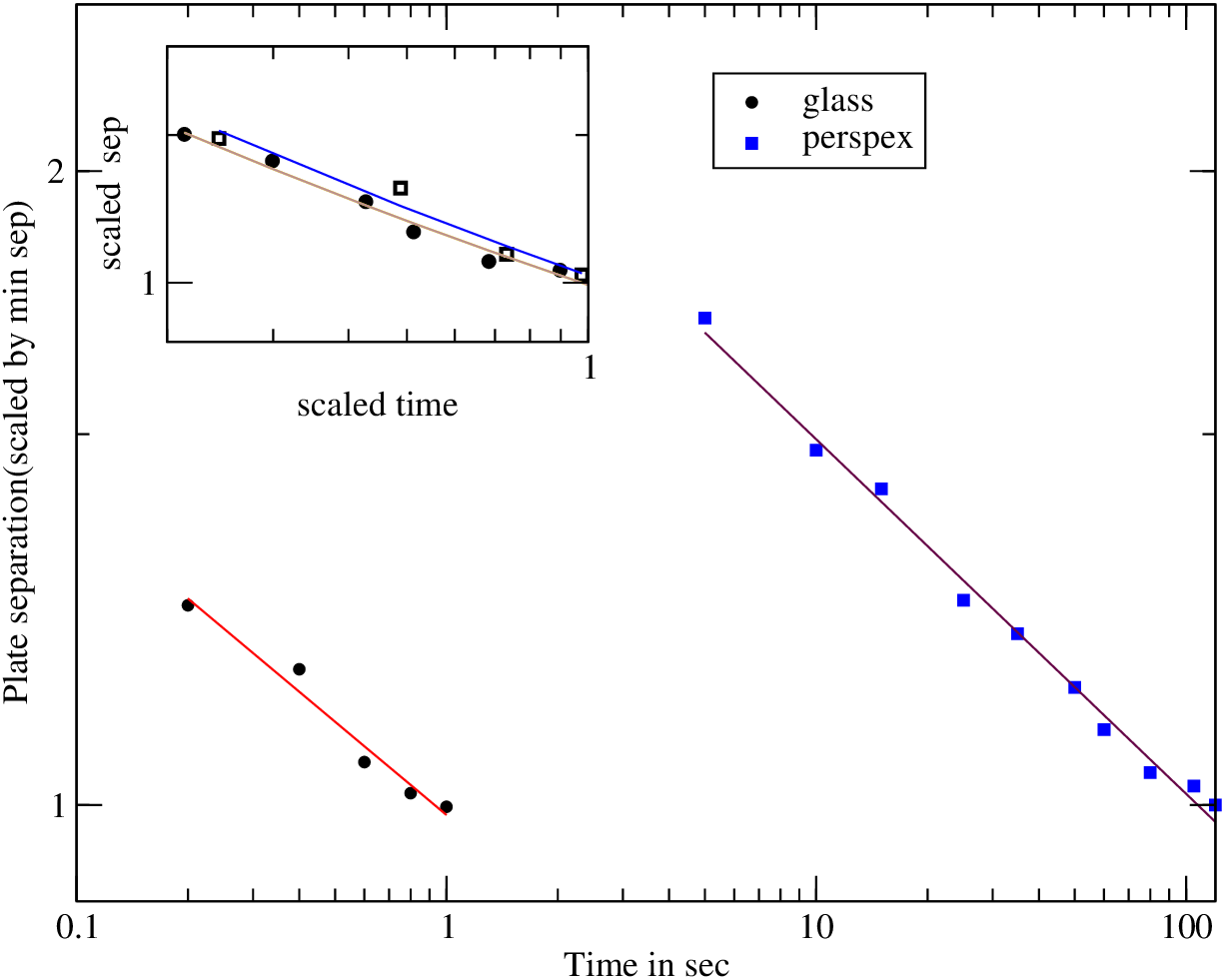}
\end{center}
\caption{Change in plate separation with time for castor oil on glass and perspex under a load of 5 kg. Inset shows same data when the time is scaled by the time for $a$ to saturate.
 } \label{psi1}
\end{figure}

The exponents for castor oil and ethylene glycol on perspex are plotted in figure(\ref{exponents}) and show a linear increase with the load.
Castor oil being less polar spreads much less on perspex compared to glass, this is shown in figure(\ref{psi1}) for a specific $M$. 

After recording the dynamics of spreading, we record the area $A_{sat}$ finally attained under a specific load. The area does not reach an absolutely constant value, but continues to increase asymptotically at a very slow rate. We note the area after an interval of 10 sec. after the load is placed. Beyond this time the increase is very little. 

\subsection{Analysis of Results}

\begin{figure}[ht]
\begin{center}
\includegraphics[width=12.0cm, angle=0]{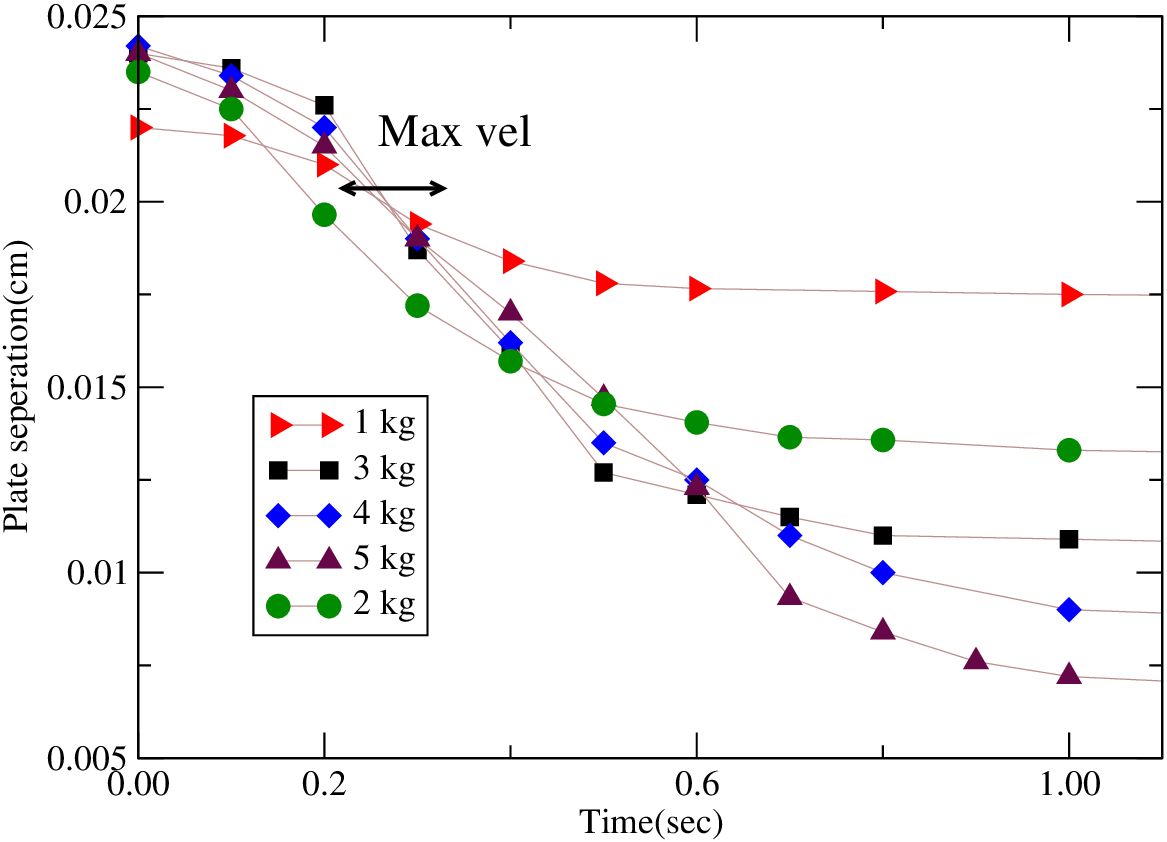}
\end{center}
\caption{Change in plate separation with time for ethylene glycol on perspex under a load of 5 kg. Data recorded at 0.1 sec intervals upto 1 sec. shows clearly the acceleration and subsequent decceleration of the plate. The approximate region where the velocity is maximum is indicated in the figure.
 } \label{fit2}
\end{figure}

What happens when a loaded plate is placed on a drop of liquid on another plate?
If the initial velocity of the plate is zero, it first gains kinetic energy falling under gravity displacing the viscous liquid. It loses this energy which is dissipated by the moving liquid. So the velocity of the plate is expected to rise initially and then fall. This is actually observed, when the video is recorded at a speed of 10 frames per second. Figure(\ref{fit2}) shows clearly that the plate separation falls at a slow rate initially, then faster up to about 0.2 seconds, then falls again to zero. We follow \cite{tabor} in calculating the final area as the load $M$ is increased. Let $M$ be the total load on the liquid film including the mass of the plate and the weights.
The pressure distribution in the liquid at any instant is shown to be
\begin{equation} p(r,t) - p(atm) = 3\mu V_f (R(t)^2-r^2)/ h(t)^3 \end{equation}

\begin{figure}[ht]
\begin{center}
\includegraphics[width=12.0cm, angle=0]{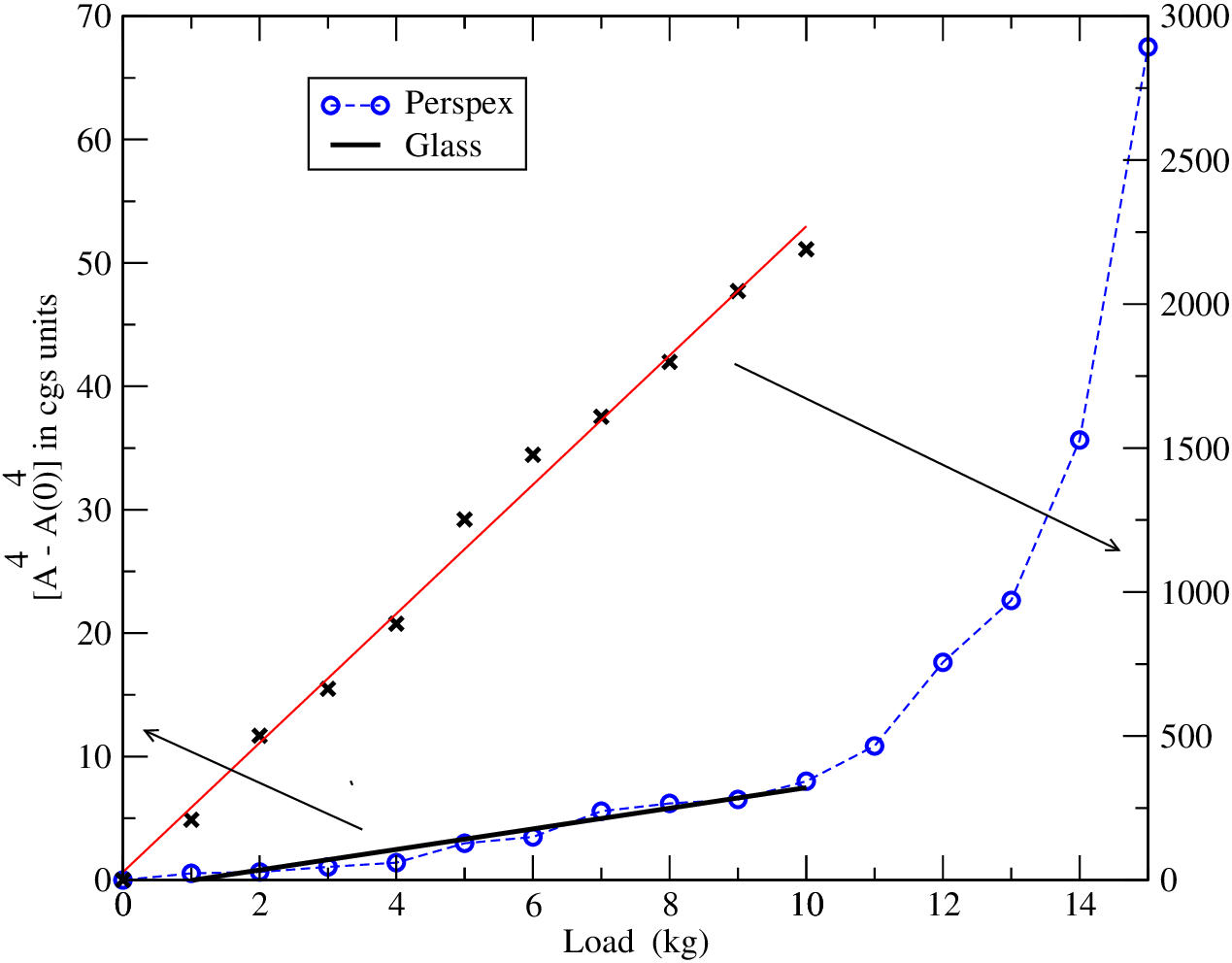}
\end{center}
\caption{$(A^4-A(0)^4)$ plotted for ethylene glycol spreading on glass and perspex as function of $M$. Upto 10 kg. load, the graphs are linear. the left hand and right y-axes correspond respectively to  perspex and glass.
 } \label{a4}
\end{figure}

Here $p(atm)$ is the atmospheric pressure outside, $\mu$ the viscosity of the liquid and $V_f$ the constant fluid volume. $R(t)$ is the instantaneous radius of the liquid blob. We consider squeezing the film under 'conserved volume' condition, in contrast to the 'constant area' condition in  \cite{tabor}. Assuming that the maximum kinetic energy corresponding to the maximum velocity $v_{max}$ attained, is dissipated fully by the viscous liquid as the plate comes to rest, we get the following relation
\begin{equation} {A_{sat}}^4 - {A_0}^{4} = 8\pi v_{max} {V_f}^2 M/3\mu  \label{area} \end{equation}
Here $\mu $ is the coefficient of viscosity of the liquid and $A_0$ is the initial area.
Graphs of ${A_{sat}}^4 - {A_0}^{4}$ against M in figure(\ref{a4})
  show that up to 10 kg. load the curve is indeed a straight line. After that there is a deviation. In the case of the glass substrate the film becomes too thin to be visible at $M>10$ kg. We have neglected the energy due to interfacial tension between the different phases and effect of the contact line, in spite of this, the agreement with experiment is pretty good. In fact for ethylene glycol on perspex, the agreement is quantitative, the calculated coefficient of $M$ in equation(\ref{area}) is 1088 gm/cm$^8$ which is quite close to the experimental slope 1492 gm/cm$^8$. However, when the substrate is glass this agreement breaks down, the calculated result here is approx 9000 gm/cm$^8$, nearly 4 orders of magnitude larger than the experimentally obtained coefficient. This indicates that since the fluid wets glass, the precursor effect is dominant here \cite{degennes} and our naive calculation fails. 
 \section{Discussion}

The variation of plate separation under different conditions is routinely studied as rheological tests for characterizing soft materials \cite{engmann}. However, in that case, the focus is on comparing the behavior of different non-Newtonian materials, rather than understanding the physics of the spreading process. Bonn \cite{bonn} review several different values of the exponent $m$, observed under different conditions. The simplest case, where the fluid is a small drop in the form of a spherical cap, balancing the viscous and capillary forces yields $m$ = 0.2 \cite{degennes,bonn}. Other values (generally higher) are reported, where inertial forces are not negligible. The effect of varying impressed loads has not been reported to our knowledge. The linear variation shown in figure(\ref{exponents}) is interesting and requires closer investigation.

The agreement of experimental $a_{sat}$ as function of $M$, with the simple analysis considering only viscous dissipation is encouraging. It will be of interest to see what leads to the deviation from this behavior at loads above 10 kg, maybe van der Waal interactions start to play a dominating role. It will also be interesting to analyze the details of the acceleration and deceleration 
of the plate as seen in figure(\ref{fit2}).

To conclude, the study of thin films under the lubrication approximation is a topic of practical importance in wetting and dewetting problems \cite{oron,engmann,sharma} and the related problems of adhesion and viscous fingering \cite{epje,creton}. 
It is also a phenomenon worth looking into from the point of view of basic science, as it may lead to a better understanding of adhesive and cohesive interactions at the microscopic level.

\section{Acknowledgment}
Authors sincerely thank Dr. Tapati Dutta for help in experiments and useful discussion. We are grateful to Prof. S P Maulik for constant encouragement and valuable suggestions. This work is supported by DST (SR/S2/CMP-22/2004) and SS is grateful to DST for grant of a research fellowhip.

\newpage

\newpage

\newpage

\newpage

\newpage

\newpage
\noindent
{\bf Figure Captions:}

\noindent
Fig.1. Plate separation with time for castor oil on perspex for different loads. Symbols represent experimental points and lines are power-law fits to the data.

\noindent
Fig.2. Exponents for the power-law fits to the area vs. time data for wthylene glycol on glass and perspex are shown as function of load. Symbols are data points and lines, linear fits.

\noindent
Fig.3. Change in plate separation with time for castor oil on glass and perspex under a load of 5 kg. Inset shows same data when the time is scaled by the time for $a$ to saturate.

\noindent
Fig.4. Change in plate separation with time for ethylene glycol on perspex under a load of 5 kg. Data recorded at 0.1 sec intervals upto 1 sec. shows clearly the acceleration and subsequent decceleration of the plate. The approximate region where the velocity is maximum is indicated in the figure.

\noindent
Fig.5. $(A^4-A(0)^4)$ plotted for ethylene glycol spreading on glass and perspex as function of $M$. Upto 10 kg. load, the graphs are linear. the left hand and right y-axes correspond respectively to  perspex and glass.

\begin{thebibliography}{99}
 \bibitem{degennes} P. G. de Gennes, Wetting: statics and dynamics, Rev. Mod. Phys. 57(1985) 827-862
 \bibitem{bonn} D. Bonn, J. Eggers, J. Meunier, E. Rolley, Wetting and spreading, to appear in Rev. Mod. Phys. 2009
 \bibitem{engmann}J. Engmann, C. Servais, A.S. Burbridge, Squeeze flow theory and applications to rheometry: A review, J. Non-Newt. Fluid Mech. 132(2005)1-27
 \bibitem{asthana}R. Asthana, N. Sobczak, Wettability,spreading and interfacial phenomena in high-temperature coatings, JOM-e 52(2000)
 \bibitem{tabor} F.R. Eirich, D. Tabor, Collisions through liquid films, Proc. Camb. Phil Soc. (1948) 566-581
 \bibitem{oron} A. Oron, S.H. Davis, S.G. Bankoff, Rev. Mod. Phys. 69(1997) 931-980
 \bibitem{sharma} A. Sharma, J. Mittal, Instability of thin liquid films by density variations: A new mechanism that mimics spinodal dewetting, Phys. Rev. Lett. 89(2002) 186101
 \bibitem{epje}S. Sinha, T. Dutta, S. Tarafdar, Adhesion and fingering in the lifting Hele-Shaw cell, Eur. Phys. J. E 25(2008) 267-275
 \bibitem{creton} J. Nase, A. Lindner, C. Creton, Pattern formation during deformation of a confined viscoelastic liquid to a soft elastic solid, Phys. Rev. Lett. 101(2008) 074503

\end{thebibliography}
\end{document}